\documentclass[12pt]{article}
\usepackage{psfig,epsfig}
\begin{document}
\def\bsig{\mbox{\boldmath $\sigma$}}                          
\def\bsig{\mbox{\boldmath $\Sigma$}}
\def\bgam{\mbox{\boldmath $\gamma$}}
\def\bgam{\mbox{\boldmath $\Gamma$}}
\def\bphi{\mbox{\boldmath $\phi$}}
\def\bphi{\mbox{\boldmath $\Phi$}}
\def\btau{\mbox{\boldmath $\tau$}}
\def\btau{\mbox{\boldmath $\Tau$}}
\def\btau{\mbox{\boldmath $\partial$}}
\def\Delc{{\Delta}_{\circ}}
\def\bp{\mid {\bf p} \mid}
\def\al{\alpha}
\def\bet{\beta}
\def\gam{\gamma}
\def\del{\delta}
\def\Del{\Delta}
\def\te{\theta}
\def\nua{{\nu}_{\alpha}}
\def\nui{{\nu}_i}
\def\nuj{{\nu}_j}
\def\nue{{\nu}_e}
\def\num{{\nu}_{\mu}}
\def\nut{{\nu}_{\tau}}
\def\2te{2{\theta}}
\def\chic#1{{\scriptscriptstyle #1}}
\def\chicl{{\chic L}}
\def\lam{\lambda}
\def\SU{SU(2)_{\chic W} \otimes U(1)_{\chic Y}}
\def\Lam{\Lambda}
\def\sig{\sigma}
\def\'#1{\ifx#1i\accent19\i\else\accent19#1\fi}
\newcommand{\be}{\begin{equation}}
\newcommand{\ee}{\end{equation}}
\newcommand{\ba}{\begin{array}}
\newcommand{\ea}{\end{array}}
\newcommand{\dis}{\displaystyle}
\newcommand{\alfad}{\frac{\dis \bar \alpha_s}{\dis \pi}}

\thispagestyle{empty}
\begin{titlepage}
\begin{center}
\hfill hep-ph/9907249\\
\hfill FTUV/99-47\\
\hfill IFIC/99-49\\
\vskip 1.0cm
{\LARGE \bf
Charge and Magnetic Moment of the Neutrino in 
the Background Field Method and in the Linear $R_{\xi}^L$ Gauge}
\end{center}
\normalsize
\vskip1cm
\begin{center}
{\large \bf L. G. Cabral-Rosetti} \footnote{E-mail: cabral@titan.ific.uv.es} 
\footnote{also I.F.I.C. Centre Mixte Universitat de Val\`encia-C.S.I.C.}
{\large \bf , J. Bernab\'eu, J. Vidal$^{\: \chic {2}}$}
\end{center}
\begin{center}
\baselineskip=13pt
{\large \it Departament de F\'{\i}sica Te\`orica\\
Universitat de Val\`encia}\\
\baselineskip=12pt
{\it E-46100 Burjassot, Val\`encia (Spain)}\\
\end{center}
\begin{center}
{\large and}
\end {center}
\begin{center}
{\large \bf A. Zepeda}
\end{center}
\begin{center}
\baselineskip=13pt
{\large \it Centro de Investigaciones y de Estudios Avanzados\\
del Instituto Polit\'ecnico Nacional (Cinvestav--I.P.N.),}\\
\baselineskip=12pt
{\it Apartado Postal 14-740, 07000 M\'exico, D.F.,  M\'exico.}\\
\vglue 0.8cm
\end{center}

\begin{abstract}

We present a computation of the charge and the magnetic moment of the 
neutrino in the recently developed electro-weak Background Field Method and 
in the linear $R_{\xi}^L$ gauge. First, we deduce a formal Ward-Takahashi 
identity which implies the immediate cancellation of the neutrino electric 
charge. This Ward-Takahashi identity is as simple as that for QED. The 
computation of the (proper and improper) one loop vertex diagrams contributing
to the neutrino electric charge is also presented in an arbitrary gauge, 
checking in this way the Ward-Takahashi identity previously obtained. Finally,
the calculation of the magnetic moment of the neutrino, in the minimal 
extension of the Standard Model with massive Dirac neutrinos, is presented, 
showing its gauge parameter and gauge structure independence explicitly.

\end{abstract}
\vfill
\end{titlepage}

\section{Introduction}

The one-loop calculation of the neutrino electric charge (NEC) and the 
neutrino magnetic moment (NMM) in the Standard Model (SM) \cite{SLG} turns 
out to be one of the simplest calculations beyond tree level and consequently 
a convenient ground where one can test methods and compare techniques.

The background field method (BFM) was first introduced by DeWitt \cite{BSD}
in the context of Quantum Gravity as a formalism for quantizing gauge
field theories while retaining explicit gauge invariance at one-loop.
The multi-loop extension of the method was given by 't Hooft \cite{GHO},
DeWitt \cite{BSW}, Boulware \cite{DGB}, Abbott \cite{LFA}, Rebhan and
Wirthumer \cite{ARG}. Using this extension of the BFM, explicit two-loop
calculations of the $\beta$ function
for pure Yang-Mills theory were made first in the Feynman gauge
by Ichinose and Omote \cite{SIM}, and later in the general  gauge
by Capper and MacLean \cite{DMC}.

The electro-weak version  of the BFM was introduced by Denner, Weiglein
and Dittmaier \cite {{ADW},{WDA}}. In this version the gauge invariance 
of the BFM effective action implies simple (QED-like)  Ward-Takahashi 
(WT) identities for the vertex function which, as a consequence, possess
desirable theoretical properties like an improved high-energy behaviour. 
The BFM gauge invariance not only admits the usual
on-shell renormalization but even simplifies its technical realization.
Moreover, the formalism provides additional advantages such as
simplifications in the Feynman rules and the possibility to use
different gauges for tree and loop lines in Feynman diagrams, allowing 
to reduce the number of graphs. 

For these reasons this paper is devoted to the presentation of the calculation
of the NEC and the NMM in the BFM as well as in the linear $R_{\xi}^L$ with 
the aim of making a useful comparison of the two methods. General expressions
carrying the full dependence in  $q^2$, masses and gauge parameter are 
necessary to see how the form factor we are interested in get rid of 
divergent (infinite) parts and gauge parameter dependences. The paper is 
devoted to the analysis of how the cancellation among the different 
contributions occur.

The paper is organized as follows. In Section 2 we get the cancellation of the
NEC by building a WT identity in the BFM and using other WT identity which is 
proved in Appendix A. In Section 3 we calculate the NEC in the BFM and in the 
linear $R_\xi^L$ gauge by an explicit calculation of the contributing Feynman 
diagrams. In Section 4 a similar calculation is presented for the NMM. This 
work is part of a most general one in the search for gauge independent form 
factors. Appendix B contains news relations between the scalar three point 
functions $C_0$ and $B_0$, useful for the calculation.

\section{Ward-Takahashi Identity in the BFM}

The most general Lorentz invariant decomposition of the $\nu \nu \gamma$
vertex is given by \cite{CWK,JEK,EAL}

\begin{equation}
{\cal M}_{\mu}\  =\  < \nu_l (p'\,) | J^{em}_\mu | \nu_l (p) >\ 
=\ i\, \bar{u}_l (p'\,) \Bigg\{F_{\chic {NEC}} (q^2, \xi)\, \gamma_\mu\,
-\, i\, \frac{F_{\chic {NMM}}(q^2, \xi)}{2 m_e}\, 
\sigma_{\mu \nu} q^{\nu} \Bigg\} u_l (p) \ , 
\end{equation}

\noindent
where $q = p - p'$, $\xi$ is the gauge parameter, $l$ refers to one of the 
leptonic families $e$, $\mu$, or $\tau$ and $F_{\chic {NEC}} (q^2, \xi)$ and 
$F_{\chic {NMM}} (q^2, \xi)$ are, respectively, the {\it Dirac} and 
{\it Pauli} form factors of the neutrino. At zero momentum transfer they
define the NEC and the NMM, respectively.

In the calculations presented in this paper, $\xi$ is the gauge parameter of 
the $W$ boson, $\xi_Q^W$ in the BFM and $\xi_W$ in the linear $R_\xi^L$ gauge.
In both cases this is the only gauge parameter in our formulas. We explicitly 
keep track of this gauge parameter in order to be able to discuss later the 
problem of defining gauge independent form factors in a more general context.

We now proceed to consider the one-loop Feynman diagrams that contribute to 
the $\nu \nu \gamma$ proper vertex. Using the Feynman rules for the BFM, 
given in Appendix A of Ref. \cite{WDA}, one immediately finds that only four 
diagrams contribute to this vertex (Figs. 1a to 4a). This is a typical 
feature of the non-linear structure of the BFM gauge fixing terms. In the 
linear $R_\xi^L$ gauge there are six proper vertex diagrams as discussed
in Ref. \cite{LRZ}. With the integral expressions for these four diagrams at 
zero momentum transfer we derive then a WT identity for the NEC following the 
analysis made in the non linear $R^{NL}_\xi$ gauge \cite{{MON},{MGG}}. The 
vertex shown in Fig. 1a can be written, in the limit of zero momentum for the 
photon, as

\begin{figure}[htb]
\begin{center}
\begin{minipage}{.4\linewidth}
\centerline{\psfig{file=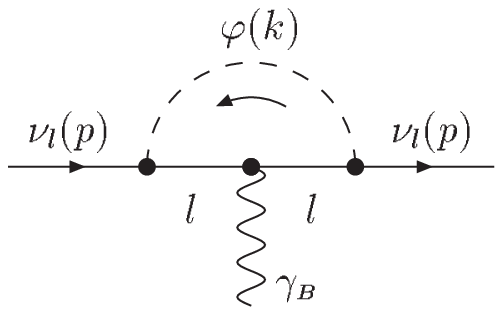,width=.9\linewidth}}
\centerline{a}
\end{minipage}
\quad
\begin{minipage}{.4\linewidth}
\centerline{\psfig{file=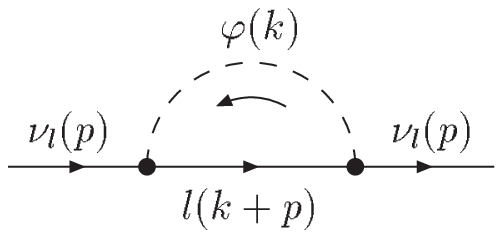,width=.9\linewidth}}
\centerline{b}
\end{minipage}
\end{center}
\caption{a) Neutrino vertex $\Lambda_\mu^{\varphi ll} (p, 0, p)$.
b) Neutrino self-energy ${\sum \nolimits}^{\varphi l} (p)$.}
\end{figure}

\begin{equation}
\Lambda_\mu^{\varphi ll} (p, 0, p)
= \frac{e^3 m^2_l}{2 s^2_W M^2_W} \int \frac{d^4 k}{(2 \pi)^4}\ 
\omega_+ S_F (k + p) \gamma_\mu S_F (k + p) 
\Delta_F (k) \omega_- \, ,
\end{equation}

\noindent
where the superscript labels the particles in the loop,
$S_F (k + p)$ and $\Delta_F (k)$ are the propagators of the fermion
and scalar field respectively, and the $\omega_\pm$ are
the chirality projectors. Let us now consider the diagram shown in 
Fig. 1b. The self-energy is given by

\begin{equation}
{\sum \nolimits}^{\varphi l} (p)
= \frac{e^2 m^2_l}{2 s^2_W M^2_W}
\int \frac{d^4 k}{(2 \pi)^4}\ \omega_+ 
\Delta_F (k) S_F (k + p) \omega_- \ .
\end{equation}

\noindent
From Eq. (2) and (3) the identity

\begin{equation}
-\, e\ \frac{\partial {\sum \nolimits}^{\varphi l} (p)}{\partial p^\mu}
= \Lambda_\mu^{\varphi ll} (p, 0, p) \, ,
\end{equation}

\noindent
follows. Similarly, the vertex of diagram 2a is given by

\begin{figure}[htb]
\begin{center}
\begin{minipage}{.4\linewidth}
\centerline{\psfig{file=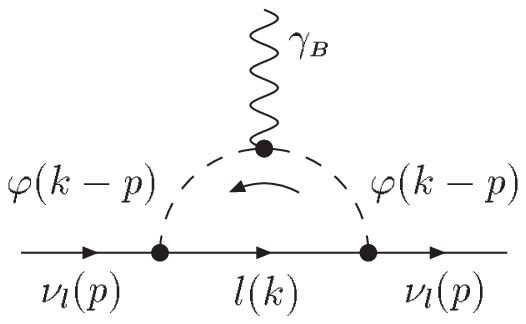,width=.9\linewidth}}
\centerline{a}
\end{minipage}
\quad
\begin{minipage}{.4\linewidth}
\centerline{\psfig{file=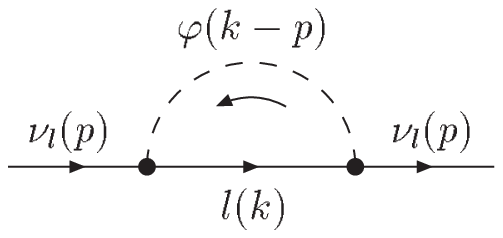,width=.9\linewidth}}
\centerline{b}
\end{minipage}
\end{center}
\caption{a) Neutrino vertex $\Lambda^{\varphi \varphi l}_\mu (p, 0, p)$.
b) Neutrino self-energy ${\sum \nolimits}^{\varphi l} (p)$.}
\end{figure}

\begin{equation}
\Lambda^{\varphi \varphi l}_\mu (p, 0, p)
= -\, \frac{e^3 m_l^2}{2 s^2_W M^2_W} \int \frac{d^4 k}{(2 \pi)^4}\ 
\omega_{+} \Delta_F (k - p) [2 (k - p)_\mu] \Delta_F
(k - p) S_F (k) \omega_{-} \, ,
\end{equation}

\noindent
and the corresponding self-energy diagram 2b is

\begin{equation}
{\sum \nolimits}^{\varphi l} (p)
= \frac{e^2 m^2}{2 s^2_W M^2_W}  
\int \frac{d^4 k}{(2 \pi)^4 }\ \omega_{+} \Delta_F (k - p)
S_F (k) \omega_{-} \, ,
\end{equation}

\noindent
so that they satisfy the identity

\begin{equation}
e\ \frac{\partial {\sum \nolimits}^{\varphi l} (p)}{ \partial p^\mu}
= \Lambda^{\varphi \varphi l}_{\mu} (p, 0, p)      \, .
\end{equation}

\noindent
Considering Eqs. (4) and  (7) we see that the contributions to the vertex 
function of diagrams 1a and 2a cancel each other. This is also the case 
in the $R_\xi^L$ gauge, where the diagrams shown in Fig. 1 and 2 (with the
change $\gamma_{\chic B} \rightarrow \gamma$) lead
to relations similar to those of equations (4) and (7).

The other two diagrams that contribute to the vertex involve a $W$ internal 
line. For the vertex shown in Fig. 3a we obtain

\begin{figure}[htb]
\begin{center}
\begin{minipage}{.4\linewidth}
\centerline{\psfig{file=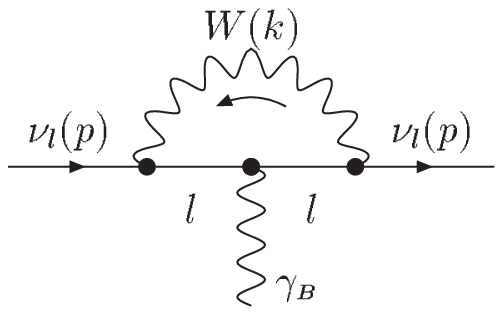,width=.9\linewidth}}
\centerline{a}
\end{minipage}
\quad
\begin{minipage}{.4\linewidth}
\centerline{\psfig{file=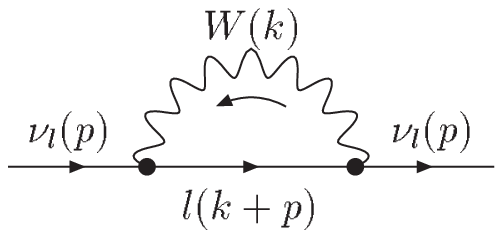,width=.9\linewidth}}
\centerline{b}
\end{minipage}
\end{center}
\caption{a) Neutrino vertex $\Lambda_\mu^{W ll} (p, 0, p)$.
b) Neutrino self-energy ${\sum \nolimits}^{W l} (p)$.}
\end{figure}

\begin{equation}
\Lambda_\mu^{W ll} (p, 0, p)
= -\, \frac{e^3}{2 s^2_W} \int \frac{d^4k}{(2 \pi)^4}\ \gamma^\alpha
\omega_{-} S_F (k + p) \gamma_\mu S_F (k + p) \gamma^\beta
D^W_{\alpha \beta} (k) \omega_{-} \, ,
\end{equation}

\noindent
where $D^W_{\alpha \beta} (k)$ is the propagator of the $W$ boson. The 
corresponding part of the self-energy of the neutrino (diagram 3b) is then

\begin{equation}
{\sum \nolimits}^{Wl} (p)
= -\, \frac{e^2}{2 s^2_W} 
\int \frac{d^4 k}{(2 \pi)^4}\ \gamma^\alpha  \omega_{-} 
D^W_{\alpha \beta} (k) \gamma^\beta S_F (k + p) \omega_{-} \, .
\end{equation}

\noindent
From Equations (8) and (9), it is straightforward to get the identity

\begin{equation}
-\, e\ \frac{\partial {\sum \nolimits}^{Wl} (p)}{\partial p^\mu} =
\Lambda^{W ll}_\mu (p, 0, p) \, ,
\end{equation}

\noindent
which is a relation similar to that obtained in the $R_\xi^L$ gauge.

The last contribution to the proper vertex, Fig. 4a, involves the non Abelian
vertex $\gamma_{\chic B} W W$. It is given by

\begin{figure}[htb]
\begin{center}
\begin{minipage}{.4\linewidth}
\centerline{\psfig{file=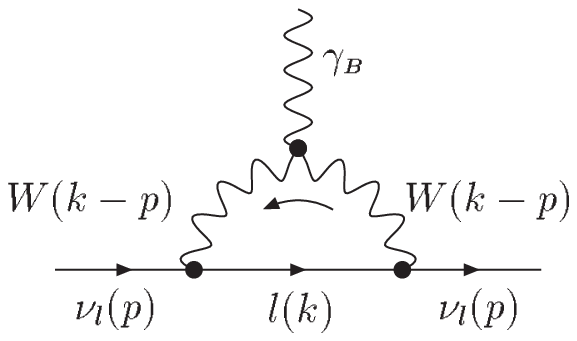,width=.9\linewidth}}
\centerline{a}
\end{minipage}
\quad
\begin{minipage}{.4\linewidth}
\centerline{\psfig{file=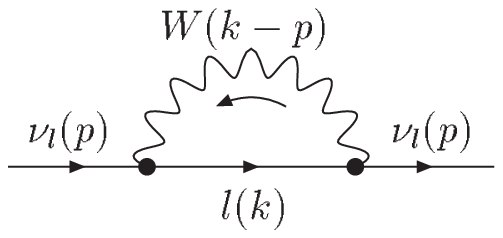,width=.9\linewidth}}
\centerline{b}
\end{minipage}
\end{center}
\caption{a) Neutrino vertex $\Lambda_\mu^{WWl} (p, 0, p)$.
b) Neutrino self-energy ${\sum \nolimits}^{W l} (p)$.}
\end{figure}

\begin{equation}
\Lambda_\mu^{WWl} (p, 0, p)
= \frac{e^2}{2 s^2_W} \int \frac{d^4 k}{(2 \pi)^4}\ \gamma^\alpha
\omega_{-} \Gamma^{W \hat{A} W}_{\alpha \beta \mu} \, 
(k - p, 0 , k - p) \gamma^\beta S_F (k) \omega_{-}  \, ,
\end{equation}

\noindent
where we have defined

\begin{equation}
\Gamma^{W \hat{A} W}_{\alpha \beta \mu} \, (k - p,0, k - p) \equiv
e D^W_{\alpha \alpha'} (k - p) V^{W \hat{A} W}_{\alpha' \beta' \mu}
[- (k - p), 0, (k - p)] D^W_{\beta \beta'} (k - p) \, ,
\end{equation}

\noindent
with $V^{W \hat{A} W}_{\alpha \beta \mu} $ being the $\hat{A} WW$-coupling
of a background field $\hat{A}$ with two W's quantum fields (its explicit
form can be found in Eqs. (A.33) and (A.34) of Ref. \cite{WDA}). Using
the {\it contracted} vertex $\Gamma^{W \hat{A} W}_{\alpha \beta \mu}$ 
defined in equation (12) it is possible to prove the WT identity

\begin{equation}
-\, e\ \frac{\partial D^W_{\alpha \beta} (l)}{\partial l^\mu} =
\Gamma^{W \hat{A} W}_{\alpha \beta \mu} (l, 0, l),
\end{equation}

\noindent 
where $l = k - p$. This formula is crucial for the computation because it
relates the quantum W's fields with the background $\hat{A}$ field and can 
not be obtained by the usual derivative procedure of the functional generator.
Equation (13) is formally the same as Eq. (5) in Ref. \cite{MON} for the non 
linear $R_\xi^{NL}$ gauge. The proof of this WT identity is given in Appendix 
A.

Taking now into account that the corresponding contribution to the 
self-energy (diagram 4b) is

\begin{equation}
{\sum \nolimits}^{Wl} (p)
= -\, \frac{e^2}{2 s^2_W} \int 
\frac{d^4 k}{(2 \pi)^4}\ \gamma^\alpha \, \omega_{-} D^W_{\alpha \beta} 
(k - p) \gamma^\beta S_F(k) \omega_{-}  \, ,
\end{equation}

\noindent
with the help of the WT identity (13), it is easy to prove the relation

\begin{equation}
e\ \frac{\partial {\sum \nolimits}^{Wl} (p)}{\partial p^\mu} = 
\Lambda^{WWl}_\mu \,(p, 0, p) \, ,
\end{equation}

\noindent
so that the contributions of diagrams 3 and 4 cancel each other again. The
relation shown in Eq. (15) doesn't exist in the linear $R_\xi^L$ gauge.

With only these four diagrams the cancellation of the NEC is not obtained in 
the $R_\xi^L$ gauge. There are two additional diagrams involving the 
$\gamma W \varphi$ vertex plus one improper diagram (transverse part of 
$\gamma Z$ self-energy) that should be considered. Only then one obtains 
\cite{LRZ} { ${\cal Q}_{\nu}^{\chic {R^L_\xi}} = 0$, in an obvious notation. 
In the BFM the last self-energy also exists, but its contribution to the NEC 
vanishes [see Eq. (34) in Ref. \cite{WDA}] because the transverse part of the 
$\gamma_{\chic B} Z_\chic {B}$ self-energy is {\it zero}.

In the BFM the four proper vertices at zero momentum transfer, satisfy the 
relation

\begin{equation}
\Lambda^{\varphi ll}_\mu + \Lambda^{\varphi \varphi l}_\mu + 
\Lambda^{W ll}_\mu + \Lambda^{W Wl}_\mu = 0 \, ,
\end{equation}

\noindent
which implies the vanishing of the NEC,

\begin{equation}
{\cal Q}_{\nu}^{\chic {BFM}}
= F_{\chic {NEC}}^{\chic {BFM}}\Big(0, \xi_Q^W\Big) = 0 \, .
\end{equation}

The proof of the exactly cancellation of the electro-magnetic proper neutrino 
vertex at one-loop, is a consequence of the most general WT identity (see 
Eq. (36) in the Ref. \cite{WDA}) valid to all orders in perturbation 
theory.

\begin{figure}[htb]
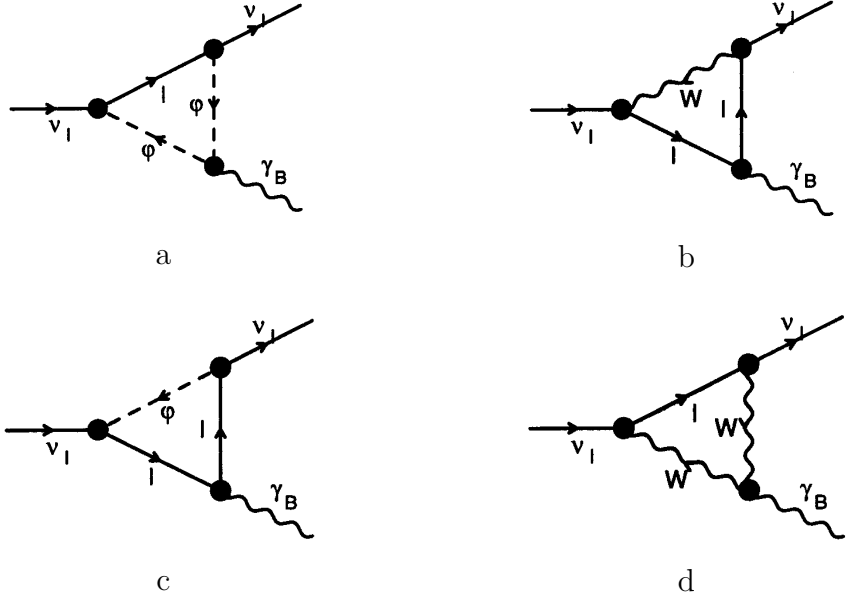

\begin{center}
\begin{minipage}{.4\linewidth}
\centerline{\psfig{file=fig5a.eps,width=.7\linewidth}}
\centerline{a}
\end{minipage}
\quad
\begin{minipage}{.4\linewidth}
\centerline{\psfig{file=fig5b.eps,width=.7\linewidth}}
\centerline{b}
\end{minipage}
\end{center}
\begin{center}
\begin{minipage}{.4\linewidth}
\centerline{\psfig{file=fig5c.eps,width=.7\linewidth}}
\centerline{c}
\end{minipage}
\quad
\begin{minipage}{.4\linewidth}
\centerline{\psfig{file=fig5d.eps,width=.7\linewidth}}
\centerline{d}
\end{minipage}
\end{center}
\caption{Vertex contributions to the NEC and NMM in the BFM.}
\end{figure}

\section{Neutrino Electric Charge, Explicit Calculation}

\noindent
In this section we present a complete calculation of the neutrino electric
charge using: a) the BFM and b) the usual formulation in the linear
$R_\xi^L$ gauge . In both cases the calculation is done for arbitrary $q^2$ 
of the photon and carrying all the gauge dependence on the $\xi$ parameter 
\cite{HES,RMF}. Cancellation of the gauge dependence, in the $q^2
\rightarrow 0$ limit, will be explicitly shown, using an expansion of the 
different contributions to the Dirac form factor around $q^2 = 0$.

\subsection{Calculation in the BFM}

\noindent 
The proper diagrams contributing to the NEC are given in Fig. 5. Using the
Feynman rules of the electroweak BFM \cite{WDA} and after some algebra one 
finds for the Dirac form factor defined in Eq. (1), at $q^2 = 0$, the 
following contributions:

\begin{equation}
\begin{array}{c}
\displaystyle
{\cal Q}_{\nu_l} \Big|_{\chic {Fig. 5a}}^{\chic {BFM}} =
F^{\chic {BFM}}_{\chic {NEC}} \Big( q^2 = 0, \xi_Q^W \Big)
\Big|_{\chic {Fig. 5a}}
= - \frac{\al e m_l^2}
{64 \pi M_W^2 s_W^2 \Big(M_W^2 - m_l^2 \xi_Q^W \Big)^2} \times 
\\[0.7cm]
\displaystyle
\Bigg\{\Big( M_W^2 - m_l^2 \xi_Q^W \Big) 
\Big(M_W^2 - 3 m_l^2 \xi_Q^W \Big)
+ 2 \Big(\xi_Q^W \Big)^2 m_l^4 B_0\, (0; m_l^2, m_l^2) 
\\[0.7cm]
\displaystyle
+ 2 M_W^2 \Big(M_W^2 - 2 m_l^2 \xi_Q^W \Big) 
B_0 \Bigg(0; \frac{M^2_W}{\xi_Q^W}, 
\frac{M^2_W}{\xi_Q^W} \Bigg) \Bigg\} \ ;
\end{array}
\end{equation}

$$
{\cal Q}_{\nu_l} \Big|_{\chic {Fig. 5b}}^{\chic {BFM}} =
F^{\chic {BFM}}_{\chic {NEC}} \Big( q^2 = 0, \xi_Q^W \Big)
\Big|_{\chic {Fig. 5b}}
= - \frac{\al e}{32 \pi (m_l^2 - M_W^2)^2 s_W^2 \xi_Q^W 
\Big(M_W^2 - m_l^2 \xi_Q^W \Big)^2} \times 
$$

$$
\Bigg\{3 M_W^2 m_l^2 \xi_Q^W \Big(M_W^2 - m_l^2 \xi_Q^W \Big)^2 
B_0\, (0; M_W^2,  M_W^2)
$$

$$
+ m_l^4 \xi_Q^W \Big[- 2 \xi_Q^W m_l^4 + M_W^2 \Big(-3 \Big(\xi_Q^W \Big)^2 
+ 4 \xi_Q^W + 3\Big) m_l^2 
+ 2 M_W^4 \Big(2 \xi_Q^W - 3\Big)\Big] B_0\, (0; m_l^2,  m_l^2)
$$

$$
- M_W^2 \Big(2 M_W^2 - m_l^2 \xi_Q^W\Big)(m_l^2 - M_W^2)^2  
B_0 \Bigg(0; \frac{M^2_W}{\xi_Q^W}, \frac{M^2_W}{\xi_Q^W} \Bigg)
$$

\begin{equation}
- \Big[- \xi_Q^W m_l^4 + M_W^2 \Big(-3 \Big(\xi_Q^W\Big)^2 
+ \xi_Q^W + 2\Big) m_l^2 + M_W^4 \Big(3 \xi_Q^W - 2\Big)\Big] 
(m_l^2 - M_W^2)\Big(M_W^2 - m_l^2 \xi_Q^W\Big) \Bigg\} \ ;
\end{equation}

\begin{equation}
{\cal Q}_{\nu_l} \Big|_{\chic {Fig. 5c}}^{\chic {BFM}}
= F^{\chic {BFM}}_{\chic {NEC}} \Big( q^2 = 0, \xi_Q^W \Big)
\Big|_{\chic {Fig. 5c}}
= - {\cal Q}_{\nu_l} \Big|_{\chic {Fig. 5a}}^{\chic {BFM}}\, ;
\end{equation}

\begin{equation}
{\cal Q}_{\nu_l} \Big|_{\chic {Fig. 5d}}^{\chic {BFM}}
= F^{\chic {BFM}}_{\chic {NEC}} \Big( q^2 = 0, \xi_Q^W \Big)
\Big|_{\chic {Fig. 5d}}
= - {\cal Q}_{\nu_l} \Big|_{\chic {Fig. 5b}}^{\chic {BFM}}\ . 
\end{equation}

\noindent
Notice that we have kept explicitly the gauge dependence in all 
equations. From Eqs. (18-21) it obviously follows that the NEC vanishes,

\begin{equation}
{\cal Q}_{\nu_l}^{\chic {BFM}} = 
{\cal Q}_{\nu_l} \Big|_{\chic {Fig. 5a}}^{\chic {BFM}} +
{\cal Q}_{\nu_l} \Big|_{\chic {Fig. 5b}}^{\chic {BFM}} +
{\cal Q}_{\nu_l} \Big|_{\chic {Fig. 5c}}^{\chic {BFM}} +
{\cal Q}_{\nu_l} \Big|_{\chic {Fig. 5d}}^{\chic {BFM}} = 0,
\end{equation}

\noindent
in agreement with the result (17), obtained using the Ward-Takahashi
identity.

Expressions (18-21) have been obtained applying a Taylor expansion, around 
$q^2 = 0$, to the complete contribution of each diagram. The complete 
contribution of {\it all} diagrams to the $F^{\chic {BFM}}_{\chic {NEC}} 
\Big(q^2, \xi_Q^W \Big)$ form factor is given in \cite{LCR}. To obtain these 
formulae we have made use of the new relations between scalar three point 
function, $C_0$, and two point functions $B_0$, given in Appendix B.

With these formulae it is easy to prove for the complete form factor
that again

\begin{equation}
{\cal Q}_{\nu_l}^{\chic {BFM}} = \lim_{q^2 \rightarrow 0}
F^{\chic {BFM}}_{\chic {NEC}} \Big(q^2, \xi_Q^W \Big) = 0 \ .
\end{equation}

\begin{figure}[htb]
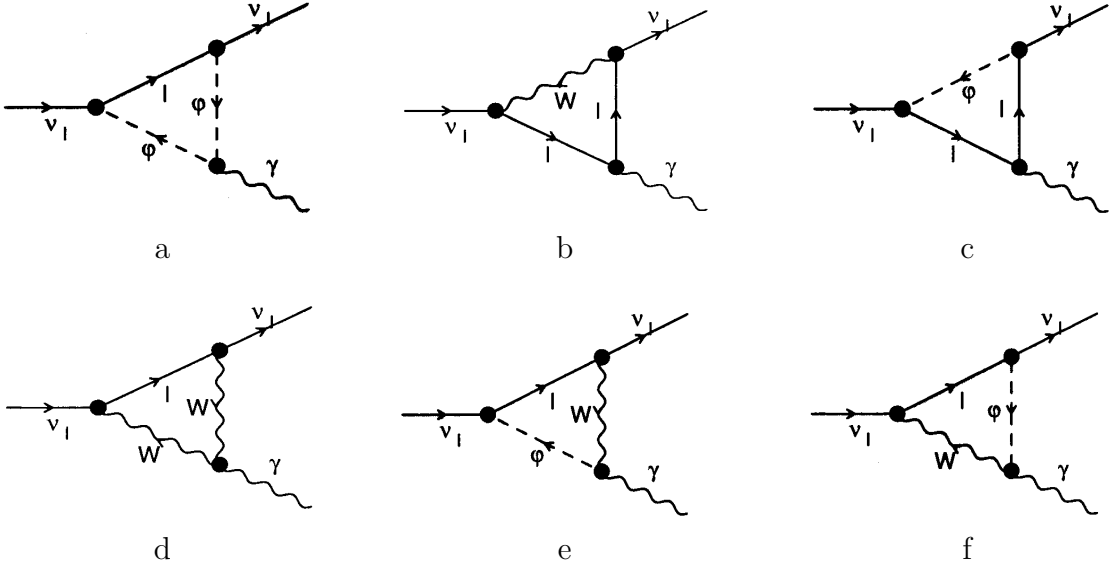

\begin{center}
\begin{minipage}{.3\linewidth}
\centerline{\psfig{file=fig6a.eps,height=3.0cm,width=4.5cm}}
\centerline{a}
\end{minipage}
\quad
\begin{minipage}{.3\linewidth}
\centerline{\psfig{file=fig6b.eps,height=3.0cm,width=4.5cm}}
\centerline{b}
\end{minipage}
\quad
\begin{minipage}{.3\linewidth}
\centerline{\psfig{file=fig6c.eps,height=3.0cm,width=4.5cm}}
\centerline{c}
\end{minipage}
\end{center}
\begin{center}
\begin{minipage}{.3\linewidth}
\centerline{\psfig{file=fig6d.eps,height=3.0cm,width=4.5cm}}
\centerline{d}
\end{minipage}
\quad
\begin{minipage}{.3\linewidth}
\centerline{\psfig{file=fig6e.eps,height=3.0cm,width=4.5cm}}
\centerline{e}
\end{minipage}
\quad
\begin{minipage}{.3\linewidth}
\centerline{\psfig{file=fig6f.eps,height=3.0cm,width=4.5cm}}
\centerline{f}
\end{minipage}
\end{center}
\caption{Vertex contributions to the NEC and NMM in the linear $R_\xi^L$ 
gauge.}
\end{figure}

\subsection{Calculation in the linear $R_\xi^L$ gauge}

\noindent
The proper vertices contributing to the NEC in the linear $R_\xi^L$ gauge are 
those given in Fig. 6. Notice that there are two diagrams that do not appear 
in the  BFM. A procedure similar to that used in the previous subsection leads
to the following result:

\begin{equation}
{\cal Q}_{\nu_l} \Big|_{\chic {Fig. 6a}}^{\chic {R_\xi^L}} =
F^{\chic {R_\xi^L}}_{\chic {NEC}} ( q^2 = 0, \xi_W )
\Big|_{\chic {Fig. 6a}}
= F^{\chic {BFM}}_{\chic {NEC}} 
\Big(q^2 = 0, \xi_Q^W \rightarrow \xi_W \Big)
\Big|_{\chic {Fig. 5a}}\ ;
\end{equation}

\begin{equation}
{\cal Q}_{\nu_l} \Big|_{\chic {Fig. 6b}}^{\chic {R_\xi^L}} =
F^{\chic {R_\xi^L}}_{\chic {NEC}} ( q^2 = 0, \xi_W )
\Big|_{\chic {Fig. 6b}}
= F^{\chic {BFM}}_{\chic {NEC}} 
\Big(q^2 = 0, \xi_Q^W \rightarrow \xi_W \Big)
\Big|_{\chic {Fig. 5b}}\ ;
\end{equation}

\begin{equation}
{\cal Q}_{\nu_l} \Big|_{\chic {Fig. 6c}}^{\chic {R_\xi^L}} =
F^{\chic {R_\xi^L}}_{\chic {NEC}} ( q^2 = 0, \xi_W )
\Big|_{\chic {Fig. 6c}}
= F^{\chic {BFM}}_{\chic {NEC}} 
\Big(q^2 = 0, \xi_Q^W \rightarrow \xi_W \Big)
\Big|_{\chic {Fig. 5c}}\ ;
\end{equation}

$$
{\cal Q}_{\nu_l} \Big|_{\chic {Fig. 6d}}^{\chic {R_\xi^L}} =
F^{\chic {R_\xi^L}}_{\chic {NEC}} ( q^2 = 0, \xi_W )
\Big|_{\chic {Fig. 6d}} =
$$

$$
\frac{\alpha e}{64 \pi (m_l^2 - M_W^2)^2 s_W^2 (M_W^2 - m_l^2 \xi_W)
(1 - \xi_W) \xi_W} \times
$$

$$
\Bigg\{ [5 m_l^2 (\xi_W + 1) + M_W^2 (\xi_W - 5)]
(m_l^2 - M_W^2) (M_W^2 - m_l^2 \xi_W)(1 - \xi_W)
$$

$$
+\, 6 [m_l^2 (1 + \xi_W) - 2 M_W^2] m_l^4 (1 - \xi_W) \xi_W 
B_0\, (0; m_l^2, m_l^2)
$$

$$
+\, 6 [m_l^2 (1 - 2 \xi_W) + M_W^2 \xi_W] M_W^2 (M_W^2 - m_l^2 \xi_W) \xi_W
B_0\, (0; M_W^2, M_W^2)
$$

\begin{equation}
-\, 6 M_W^2 (m_l^2 - M_W^2)^2
B_0 \Bigg(0; \frac{M^2_W}{\xi_W}, \frac{M^2_W}{\xi_W} \Bigg) \Bigg\}\ ;
\end{equation}

$$
{\cal Q}_{\nu_l} \Big|_{\chic {Fig. 6e}}^{\chic {R_\xi^L}} =
F^{\chic {R_\xi^L}}_{\chic {NEC}} ( q^2 = 0, \xi_W )
\Big|_{\chic {Fig. 6e}} =
$$

$$
{\cal Q}_{\nu_l} \Big|_{\chic {Fig. 6f}}^{\chic {R_\xi^L}} =
F^{\chic {R_\xi^L}}_{\chic {NEC}} ( q^2 = 0, \xi_W )
\Big|_{\chic {Fig. 6f}}
$$

$$
= \frac{\alpha e}{64 \pi (m_l^2 - M_W^2) s_W^2 (M_W^2 - m_l^2 \xi_W)^2
(1 - \xi_W)} \times
$$

$$
\Bigg\{ -\, m_l^2 (m_l^2 - M_W^2) (M_W^2 - m_l^2 \xi_W) 
(1 - \xi_W)
$$

$$
+\, \xi_W [m_l^2 (3 \xi_W + 1) - 4 M_W^2] m_l^4 (1 - \xi_W)
B_0\, (0; m_l^2, m_l^2) 
$$

$$
+ 3 \xi_W  m_l^2 (M_W^2 - m_l^2 \xi_W)^2 
B_0 (0; M^2_W, M^2_W)
$$

\begin{equation}
+\, \xi_W [m_l^2 (2 \xi_W + 1) - 3 M_W^2] m_l^2 (m_l^2 - M_W^2)
B_0 \Bigg(0; \frac{M^2_W}{\xi_W}, \frac{M^2_W}{\xi_W} \Bigg)\Bigg\}\ .
\end{equation}

\noindent
Therefore, as in the BFM, the contributions from the diagrams 6a and 6c cancel
each other, but the sum of the contributions of the diagrams 6b, 6d, 6e and 6f
does not vanish. However, there exists an important relation between BFM and 
$R_\xi^L$ gauge

\begin{equation}
\begin{array}{c}
\displaystyle
F^{\chic {BFM}}_{\chic {NEC}} 
\Big(q^2 = 0, \xi_Q^W \rightarrow \xi_W \Big)
\Big|_{\chic {Fig. 5d}}
= F^{\chic {R_\xi^L}}_{\chic {NEC}} ( q^2 = 0, \xi_W )
\Big|_{\chic {Fig. 6d}}
\\[0.7cm]
\displaystyle
+\, F^{\chic {R_\xi^L}}_{\chic {NEC}} ( q^2 = 0, \xi_W )
\Big|_{\chic {Fig. 6e}}
+\,  F^{\chic {R_\xi^L}}_{\chic {NEC}} ( q^2 = 0, \xi_W )
\Big|_{\chic {Fig. 6f}} + {\sum \nolimits}_{\gamma Z}^{\chic {R_\xi^L}}
\end{array}
\end{equation}

\begin{figure}[htb]
\centerline{\psfig{file=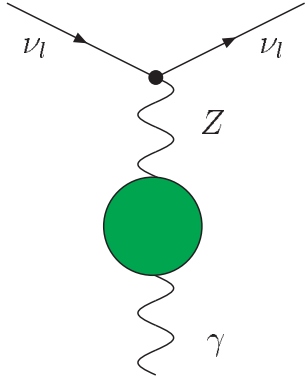,width=.25\linewidth}}
\caption{Improper vertex contribution to the NEC in the linear 
$R_\xi^L$ gauge.}
\end{figure}

\noindent where ${\sum \nolimits}_{\gamma Z}^{\chic {R_\xi^L}}$ is the 
bosonic contribution of the transverse part of the $\gamma Z$ self-energy 
(improper vertex) \cite{LRZ,ADF}, shown in Fig.~7 . This self-energy is 
computed in \cite{LCR} for arbitrary $q^2$ and, from there, we obtain

\begin{equation}
\begin{array}{c}
\displaystyle
{\sum \nolimits}_{\gamma Z}^{\chic {R_\xi^L}} = 
\frac{e {\sum \nolimits}_T^{\gamma Z} (q^2 = 0, \xi_W)}{4 s_W c_W M_Z^2} 
= \frac{\alpha e}{64 \pi s^2_W \xi_W (1 -\xi_W)} \times
\\[0.7cm]
\displaystyle 
\Bigg\{(5 \xi_W + 1)(1 -\xi_W) - 6 \xi_W^2 B_0 (0; M^2_W, M_W^2)
+ 2 (2 \xi_W + 1) B_0 \Bigg(0; \frac{M^2_W}{\xi_W}, \frac{M^2_W}{\xi_W} 
\Bigg) \Bigg\}\ .
\end{array}
\end{equation}

Summing up all these contributions one get an equation for the NEC which 
is formally the same as that given in Eq. (22), so that 

\begin{equation}
\begin{array}{c}
\displaystyle
{\cal Q}_{\nu_l}^{\chic {R_\xi^L}} = 
{\cal Q}_{\nu_l} \Big|_{\chic {Fig. 6a}}^{\chic {R_\xi^L}} +
{\cal Q}_{\nu_l} \Big|_{\chic {Fig. 6b}}^{\chic {R_\xi^L}} +
{\cal Q}_{\nu_l} \Big|_{\chic {Fig. 6c}}^{\chic {R_\xi^L}}
\\[0.7cm]
\displaystyle
+ {\cal Q}_{\nu_l} \Big|_{\chic {Fig. 6d}}^{\chic {R_\xi^L}} +
{\cal Q}_{\nu_l} \Big|_{\chic {Fig. 6e}}^{\chic {R_\xi^L}} +
{\cal Q}_{\nu_l} \Big|_{\chic {Fig. 6f}}^{\chic {R_\xi^L}} + 
{\sum \nolimits}_{\gamma Z}^{\chic {R_\xi^L}} = 0\ .
\end{array}
\end{equation}

\section{Neutrino Magnetic Moment}

\subsection{Computation in the BFM and the linear $R_\xi^L$ gauge}

\noindent 
In the {\it minimal} extension of the Standard Model (SM), in which a
right handed neutrino for each family is added, a neutrino magnetic moment 
(NMM) arises naturally. We shall compute it here, both in the BFM and the 
$R_\xi^L$ gauge, considering massive Dirac neutrinos with no flavor mixing 
\cite{PAL}. It is given by the Pauli form factor of Eq. (1),

\begin{equation}
{\mu}_{\nu_l} = \lim_{q^2 \rightarrow 0}
F^{\chic {BFM ; R_\xi^L}}_{\chic {NMM}} (q^2, m_{\nu_l}, \xi)\ .
\end{equation}

As in the previous section the contributing diagrams are given in 
Fig. 5 for the BFM and in Fig. 6 for the linear $R_\xi^L$ gauge. 
We thus obtain

\begin{equation}
\mu_{\nu_l} \big|_{\chic {Fig. 5a}}^{\chic {BFM}} =
\mu_{\nu_l} \Big|_{\chic {Fig. 6a}}^{\chic {R_{\xi}^L}}
= \frac{e m_{\nu_l} G_{\chic F}}{4 \pi^2 \sqrt{2}}
\Bigg\{ \frac{5 {\xi}}{12} x + \cdots \Bigg\} \ , 
\end{equation}

\begin{equation}
\mu_{\nu_l} \big|_{\chic {Fig. 5b}}^{\chic {BFM}} =
\mu_{\nu_l} \Big|_{\chic {Fig. 6b}}^{\chic {R_{\xi}^L}}
= \frac{e m_{\nu_l} G_{\chic F}}{4 \pi^2 \sqrt{2}}
\Bigg\{\frac{2}{3} + x \Bigg[ \frac{5 {\xi}}{3} 
- \frac{1}{2} + {\xi} \Big(\log x 
+ \log {\xi} \Big)\Bigg] + \cdots \Bigg\} \ , 
\end{equation}

\begin{equation}
\mu_{\nu_l} \big|_{\chic {Fig. 5c}}^{\chic {BFM}} =
\mu_{\nu_l} \Big|_{\chic {Fig. 6c}}^{\chic {R_{\xi}^L}}
= \frac{e m_{\nu_l} G_{\chic F}}{4 \pi^2 \sqrt{2}}
\Bigg\{- x {\xi} \Bigg[ \frac{5}{3} + \log x 
+ \log {\xi} \Bigg] + \cdots \Bigg\} \ , 
\end{equation}

\begin{equation}
\mu_{\nu_l} \Big|_{\chic {Fig. 6d}}^{\chic {R_{\xi}^L}}
= \frac{e m_{\nu_l} G_{\chic F}}{4 \pi^2 \sqrt{2}}
\Bigg\{ \frac{1}{12} \Big(7 - 5 x {\xi} \Big) 
+ \cdots \Bigg\} \ , 
\end{equation}

\begin{equation}
\mu_{\nu_l} \Big|_{\chic {Fig. 6e}}^{\chic {R_{\xi}^L}}
= \mu_{\nu_l} \Big|_{\chic {Fig. 6f}}^{\chic {R_{\xi}^L}}
= \frac{e m_{\nu_l} G_{\chic F}}{4 \pi^2 \sqrt{2}}
\Bigg\{ \frac{1}{8} \Big(1 - x \Big) + \cdots \Bigg\} \ , 
\end{equation}

\begin{equation}
\mu_{\nu_l} \big|_{\chic {Fig. 5d}}^{\chic {BFM}} =
\mu_{\nu_l} \Big|_{\chic {Fig. 6d}}^{\chic {R_{\xi}^L}} +
\mu_{\nu_l} \Big|_{\chic {Figs. 6e + 6f}}^{\chic {R_{\xi}^L}}
= \frac{e m_{\nu_l} G_{\chic F}}{4 \pi^2 \sqrt{2}}
\Bigg\{ \frac{5}{6} - x \Bigg[ \frac{1}{4} 
+ \frac{5 {\xi}}{12} \Bigg] + \cdots \Bigg\} \ , 
\end{equation}

\noindent 
where $G_F$ is the Fermi constant,
$x = m_l^2/M^2_W$ and the expressions are given up to second order
in the $x$ expansion. As we can see, the contributions from diagrams 5a, 5b 
and 5c are the same in the BFM and in the linear $R_\xi^L$ gauge (just by 
changing $\xi_Q^W \rightarrow \xi_W$), and the sum of the contributions of 
diagrams 6d, 6e and 6f is equal to the contribution of figure 5d (with the 
same substitution in the gauge parameter). The dependence in the gauge 
parameter $\xi$ is explicitly shown in the formulae. The cancellation of 
the gauge parameter is then obtained after summing up all contributions. 
However, to leading (first) order in $m_\nu$, each diagram separately is 
finite, as well as gauge parameter and gauge structure (linear or not linear)
independent. In this limit we recover the well know expression of the neutrino
magnetic moment (NMM) \cite{LES}

\begin{equation}
\mu_{\nu_l} = \frac{3 e m_{\nu_l} G_{\chic F}}{8 \pi^2 \sqrt{2}}\ .
\end{equation}

\noindent
Even in the $R_\xi^L$ gauge, there are no contributions to the magnetic 
anomaly coming from the $\gamma Z$ self-energy. 

The exact expressions for the form factor 
$F_{\chic {NMM}}^{\chic{BFM ;R_{\xi}^L}}(m_{\nu_l}, {\xi})$ (at $q^2 = 0$) 
to all orders in x is given in \cite{LCR}. From there and using the relations 
of Appendix B one gets the exact expression

\begin{equation}
\begin{array}{c}
\displaystyle
F_{\chic {NMM}}^{\chic {BFM;R_{\xi}^L}}(m_{\nu_l})
= \frac{\alpha e}{4 \pi} \frac{3 m_{\nu_l}}{4 s^2_W (m_l^2 - M_W^2)^2}
\times
\\[0.7cm]
\displaystyle
\Bigg \{ \frac{m_l^4 - 5 m_l^2 M_W^4 + 2 M_W^4}{2 M_W^2}
+ \frac{m_l^4 [B_0 (0; M^2_W, M^2_W) - B_0\, (0; m_l^2,m_l^2)]}
{m_l^2 - M_W^2}\Bigg \} \ ,
\end{array}
\end{equation}

\noindent
where the cancellation of the gauge dependence has been made evident. 
Using now the expression of the scalar two point function, $B_0$, one 
gets \cite{PAL}

\begin{equation}
F_{\chic {NMM}}^{\chic {BFM;R_{\xi}^L}}(m_{\nu_l})
= - \frac{e m_{\nu_l} G_{\chic F}}{4 \pi^2 \sqrt{2}}
\Bigg \{ \frac{3[x^3 + 2(\log x - 3)x^2 + 7x -2]}{4(1-x)^3} 
\Bigg\}\ ,
\end{equation}

\noindent
that at leading order in $x$ coincides with Eq. (39).

\section{Conclusions}

\noindent 
We have calculated the neutrino electric charge in the background field
method and in the linear $R_\xi^L$ gauge for arbitrary values of the gauge 
fixing parameters. In the BFM it has been obtained in two different ways: 
a) making use of the usual Ward-Takahashi identities for self-energy and 
vertex one loop diagrams in addition to a new one, deduced for the non 
Abelian vertex of a background field $\hat A$ and two quantum W's 
fields ($W {\hat A} W$); and b) by computing explicitly the contributing 
diagrams and checking in this way the WT identities.
In the linear $R_\xi^L$ gauge more diagrams have to be taken into account,
including the improper vertex. Therefore the BFM enjoys the simplicity of 
theories with nonlinear gauge fixing terms: fewer diagrams and simple WT 
identities. We have found that, as expected, both methods lead to the same 
result, namely: the neutrino electric charge vanishes.

For Dirac massive neutrinos, we have calculated the magnetic moment in both 
BFM and $R_\xi^L$ gauge, reproducing known results at leading order in 
$(\frac{m_1}{M_W})$. We have established the diagrammatic correspondence 
between the two methods and the gauge parameter cancellations. Finally, we
showed its gauge parameter and gauge structure independence explicitly.

\section*{Acknowledgments}

\noindent
L.G.C.R. would like to thank T. Hahn, S. Dittmaier and G. Weiglein
for helpful comments about FeynArts 2.0, BFM and Ph.D. Thesis respectively; 
M.C. Gonz\'alez-Garc\'{\i}a, M.A. Sanchis-Lozano, A. P\'erez-Lorenzana and S. 
Pastor for many discussions; and especially R. Mertig for introducing him to 
FeynCalc 3.0 and for his hospitality at NIKHEF. He is supported by a 
fellowship from the Direcci\'on General de Asuntos del Personal Acad\'emico 
de la Universidad Nacional Aut\'onoma de M\'exico (D.G.A.P.A.-U.N.A.M.). This 
work was supported in part by CICYT, under Grant AEN-96/1718, Spain. A.Z. 
acknowledges the hospitality of the Theory Group of the Fermi National 
Laboratory.

\section*{Appendix A. Ward Takahashi identity with a \\
background photon and two W's quantum fields}

\noindent
When the momentum of the background photon is zero, the vertex can
be written as [see Eqs. (A.33) \footnote{NOTE: We make the change
$\xi_{Q}^W \rightarrow 1/\xi_{Q}^W$ in the original Feynman \cite{WDA}
rules for future applications.} from Ref. \cite{WDA}]

$$
V_{\alpha' \beta' \mu}^{W \hat{A} W} (- l, 0, l) =
- ie \Big[- 2l_\mu g_{\alpha' \beta'} 
+ \Big(1 - \xi_Q^W \Big) g_{\mu \alpha'} 
l_{\beta'} + \Big(1 - \xi_Q^W \Big) g_{\beta' \mu} l_{\alpha'} \Big] \, .\\
\eqno{(A.1)}
$$

\noindent
Contracting the vertex tensor (A.1) with the $W's$ quantum propagators
we get

$$
\Gamma_{\alpha \beta \mu}^{W \hat{A} W} (l, 0, l) = e \,
\left\{ \frac{\Big(1 - \xi_Q^W \Big) \, (g_{\mu \alpha} l_\beta + g_{\beta \mu}
l_\alpha) - 2 g_{\alpha \beta} l_\mu}{\Big(l^2 - M^2_W \Big)^2} \right. 
$$

$$
- \frac{\Big(1 - \xi_Q^W \Big) \theta l^2 (g_{\alpha \mu} 
l_\beta + g_{\mu \beta} 
l_\alpha)}{\Big(l^2 - M^2_W \Big)^2 \Big(l^2 - \frac{M^2_W}{\xi_Q^W}\Big)} +
\frac{4 \theta l_\alpha l_\beta l_\mu}{
\Big(l^2 - M^2_W \Big)^2 \Big(l^2 - \frac{M^2_W}{\xi_Q^W}\Big)}
$$

$$
\left. - \frac{2 \Big (1 - \xi_Q^W \Big) \theta l_\alpha l_\beta l_\mu}{
\Big(l^2 - M_W^2 \Big)^2
\Big(l^2 - \frac{M^2_W}{\xi_Q^W}\Big)} - \frac{2 \xi_Q^W \theta^2 l^2 
l_\alpha l_\beta l_\mu}{\Big(l^2 - M_W^2 \Big)^2
\Big(l^2 -   \frac{M^2_W}{\xi_Q^W}\Big)^2} \right\},
\eqno{(A.2)}
$$

\noindent
where we have defined

$$
\theta \equiv 1 - \frac{1}{\xi_Q^W}. \\
\eqno{(A.3)}
$$

\noindent
On the other hand, the derivative of the boson propagator $W$ is

$$
- e \, \frac{\partial D^W_{\alpha \beta} (l)}{\partial
l^\mu} = e \, \left\{ \frac{- \theta (l_\alpha g_{\mu \beta}
+ l_\beta g_{\mu \alpha})}{\Big(l^2 - M^2_W \Big) 
\Big(l^2 - \frac{M^2_W}{\xi_Q^W}\Big)} 
- \frac{2 l_\mu g_{\alpha \beta}}{\Big(l^2 - M^2_W \Big)^2}
\right.
$$

$$
+ \left. \frac{2 \theta l_\alpha l_\beta l_\mu}{(l^2
- M^2_W) \Big(l^2 - \frac{M^2_W}{\xi_Q^W}\Big)} \left[ \frac{1}{
\Big(l^2 - M^2_W \Big)}
+ \frac{1}{\Big(l^2 - \frac{M^2_W}{\xi_Q^W}\Big) } \right] \right\} \, , \\
\eqno{(A.4)}
$$

\noindent
so that the comparison of (A.2) and (A.4) gives

$$
- e \, \frac{\partial D^W_{\alpha \beta} (l)}{\partial
l^\mu} = \Gamma^{W \hat{A} W}_{\alpha \beta \mu} (l, 0, l)
\, . \\
\eqno{(A.5)}
$$

\section*{Appendix B. Scalar one loop integrals.}

We present in this Appendix all the scalar one-loop integrals that
we encountered in the calculation of the NEC and NMM \cite{GHV,MRA}.
The one-loop integrals $A_0, B_0$ and $C_0$ are not independent in 
special kinematical situations \cite{ADF,MBH,WHP,JFF}.

$\bullet$ {\it One-point function}:

$$
A_0 (m^2) = m^2 \Bigg(\Delta - \log \frac{m^2}{\mu^2 } + 1 \Bigg)\ \  
with \ \ 
\Delta = \frac{2}{\epsilon} - \gamma_E + \log 4 \pi\ , \\
\eqno{(B.1)}
$$

\noindent
\ \ \ \ \ \ \ \ and where $\gamma_E$ is the Euler-Mascheroni's constant.

$\bullet$ {\it Two-point functions}:

$$
B_0 (0; m^2, m^2) = \Delta -  \log \frac{m^2}{\mu^2}\ \ ;\ \ 
B_0 (0; 0, m^2) = B_0 (0; m^2, m^2) + 1 \quad;
$$

$$
B_0 (0; 0, m^2) = B_0 (m^2; 0, m^2) - 1 \ \ \ ;\ \ \quad A_0 (m^2) =
m^2 B_0 (0; 0, m^2)\  ;
$$

$$
B_0 (m^2; 0, m^2) = B_0 (0; m^2, m^2) + 2 \ \ ;\ \ 
A_0  (m^2) = m^2 [1 + B_0 (0; m^2, m^2)] \ ;
$$

$$
(m_1^2 - m_2^2) B_0 (0; m_1^2, m_2^2) = A_0 (m_1^2) - A_0 (m_2^2)\, ;
$$

$$
B_0(q^2; m_1^2, m_2^2) = B_0(0; m_1^2, m_2^2)
$$

$$
\, + \frac{q^2}{2 (m_1^2 - m_2^2)^2} \Big\{ 
(m_1^2 + m_2^2)[1 + B_0(0, m_1^2, m_2^2)] 
- [A_0(m_1^2) + A_0(m_2^2)] \Big\} 
$$

$$
+\, \frac{q^4}{6 (m_1^2 - m_2^2)^4}\Bigg\{
m_1^4 + 10 m_1^2 m_2^2 + m_2^4 
$$

$$
+\, 3 (m_1^2 + m_2^2)
\Big\{(m_1^2 + m_2^2) B_0(0; m_1^2, m_2^2) 
- [A_0(m_1^2) + A_0(m_2^2)]\Big \} \Bigg\} + {\cal O} (q^6) \ , \\
\eqno{(B.2)}
$$

$\bullet$  {\it Three point functions} \cite{MAS}:

$$
C_0 (0, q^2, 0; m^2, M^2, M^2)
= \frac{1}{m^2 - M^2} \Big\{ B_0 (0; M^2, m^2) - B_0 (0; M^2, M^2) \Big\}
$$

$$
-\, \frac{q^2}{12 M^2 (m^2 - M^2)^4 } \Big\{ 
(m^2 - M^2)(2 m^4 + 5 M^2 m^2 - M^4)
$$

$$
+\, 6 M^2 m^4 \Big[ B_0 (0; m^2, m^2)
- B_0 (0; M^2, M^2)\Big] \Big\}
$$

$$
-\, \frac{q^4}{180 M^4 (m^2 - M^2)^6} \Big\{ (m^2 - M^2)(3 m^8 - 27 M^2 m^6
- 47 M^4 m^4 + 13 M^6 m^2 - 2 M^8)
$$

$$
-\, 60 M^4 m^6 \Big[ B_0 (0; m^2, m^2) - B_0 (0; M^2, M^2) \Big] \Big\}
+ {\cal O} (q^6) \, ;
$$

$$
C_0 (r^2, 0, r^2; m^2, M^2, M^2)
= \frac{1}{(m^2 - M^2)^2} \Big\{(m^2 - M^2) + 
m^2 \Big( B_0 (0; m^2, m^2) - B_0 (0; M^2, M^2) \Big) \Big\}
$$

$$
+ \frac{r^2}{2 (m^2 - M^2)^4 } \Big\{ 5 m^4 - 4 M^2 m^2 - M^4 
+ 2 (m^4 + 2 M^2 m^2) \Big[ B_0 (0; m^2, m^2) 
- B_0 (0; M^2, M^2)\Big] \Big\}
$$

$$
+ \frac{r^4}{3 (m^2 - M^2)^6} \Big\{10 m^6 + 9 M^2 m^4 
- 18 M^4 m^2 - M^6 
$$

$$
+ 3 (m^6 + 6 M^2 m^4 + 3 M^4 m^2)\Big[ 
B_0 (0; m^2, m^2) - B_0(0; M^2, M^2)  \Big] \Big\} 
+ {\cal O} (r^6) \, ;
$$

$$
C_0 (0, q^2, 0; m_1^2, m_2^2, m_3^2)
= \frac{1}{m_2^2 - m_1^2} \Big\{ B_0 (0; m_3^2, m_2^2) 
- B_0 (0;m_3^2, m_1^2) \Big\}
$$

$$
-\, \frac{q^2}{2}  \Bigg\{ \frac{B_0 (0; m_1^2, m_1^2) m_1^4}
{(m_1^2 - m_2)^2 (m_1^2 - m_3^2)^2}
+  \frac{m_1^2 m_2^2 - 2 m_3^2 m_2^2 + m_1^2 m_3^2}
{(m_1^2 - m_2^2) (m_1^2 - m_3^2) (m_2^2 - m_3^2)^2}
$$

$$
-\, \frac{B_0 (0; m_2^2, m_2^2) m_2^2 (m_2^4 + m_3^2 m_2^2 - 2 m_1^2 m_3^2)}
{(m_1^2 - m_2^2)^2 (m_2^2 - m_3^2)^3} 
- \frac{B_0 (0; m_3^2, m_3^2) m_3^2 (- m_3^4 - m_2^2 m_3^2 + 2 m_1^2 m_2^2)}
{(m_1^2 - m_3^2)^2 (m_2^2 - m_3^2)^3} \Bigg\}
$$

$$
+\, \frac{q^4}{3} \Bigg\{ \frac{B_0 (0; m_1^2, m_1^2) m_1^6}
{(m_1^2 - m_2^2)^3 (m_1^2 - m_3^2)^3} - \frac{B_0 (0; m_2^2, m_2^2) m_2^2}
{(m_1^2 - m_2^2)^3 (m_2^2 - m_3^2)^5} \Big[ m_2^8 + (3 m_1^4 - 9 m_2^2 m_1^2
+ 4 m_2^4) m_3^2 m_2^2
$$

$$
+\, (3 m_1^4 - 3 m_2^2 m_1^2 + m_2^4) m_3^4 \Big] 
+ \frac{B_0 (0; m_3^2, m_3^2) m_3^2}
{(m_1^2 - m_3^2)^3 (m_2^2 - m_3^2)^5} \Big[ 3 m_2^2 (m_2^2 + m_3^2) m_1^4
- 3 m_2^2 m_3^2 (m_2^2 + 3 m_3^2) m_1^2
$$

$$
+\, m_3^4 (m_2^4 + 4 m_3^2 m_2^2 + m_3^4) \Big] 
- \frac{1}{2 (m_1^2 - m_2^2)^2 (m_1^2 - m_3^2)^2 (m_2^2 - m_3^2)^4}
\Big[ (m_2^4 + 10 m_3^2 m_2^2 + m_3^4) m_1^6
$$

$$
-\, 3 (m_2^2 + m_3^2)(m_2^4 + 4 m_3^2 m_2^2 + m_3^4)m_1^4
+ m_3^2(m_2^6 + 40 m_3^2 m_2^4 - 11 m_3^4 m_2^2 + 6 m_3^6)m_1^2
$$

$$
-2 m_3^6( 10 m_2^4 - 5m_3^2 m_2^2 + m_3^4) \Big] \Bigg\} + {\cal O} (q^6) \, ;
$$

$$
C_0 (r^2, 0, r^2; m_1^2, m_2^2, m_3^2)
= \frac{m_1^2 B_0 (0; m_1^2, m_1^2)}{(m_1^2 - m_2^2)(m_1^2 - m_3^2)}
- \frac{m_2^2 B_0 (0; m_2^2, m_2^2)}{(m_1^2 - m_2^2)(m_2^2 - m_3^2)}
+ \frac{m_3^2 B_0 (0; m_3^2, m_3^2)}{(m_1^2 - m_3^2)(m_2^2 - m_3^2)}
$$

$$
+\, r^2 \Bigg\{ \frac{3 m_1^4 - m_2^2 m_1^2 - m_3^2 m_1^2 - m_2^2 m_3^2}
{2 (m_1^2 - m_2^2)^2 (m_1^2 - m_3^2)^2}
+ \frac{m_1^2}{(m_2^2 - m_3^2)} \Bigg[ \frac{m_3^2 B_0 (0; m_3^2, m_3^2)}
{(m_1^2 - m_3^2)^3} - \frac{m_2^2 B_0 (0; m_2^2, m_2^2)}
{(m_1^2 - m_2^2)^3} \Bigg]
$$

$$
+ \frac{m_1^2 \Big( m_1^6 - 3 m_2^2 m_3^2 m_1^2 + m_2^2 m_3^4 
+ m_2^4 m_3^2 \Big) B_0 (0; m_1^2, m_1^2) }
{(m_1^2 - m_2^2)^3 (m_1^2 - m_3^2)^3} \Bigg\}
$$

$$
+\, r^4 \Bigg\{ \frac{m_1^2}{(m_2^2 - m_3^2)} \Bigg[
\frac{m_3^2 (m_1^2 + m_3^2) B_0 (0; m_3^2, m_3^2)}{(m_1^2 - m_3^2)^5}
- \frac{m_2^2 (m_1^2 + m_2^2) B_0 (0; m_2^2, m_2^2)}{(m_1^2 - m_2^2)^5}
\Bigg]
$$

$$
+\, \frac{m_1^2 B_0 (0; m_1^2, m_1^2)}
{(m_1^2 - m_2^2)^5 (m_1^2 - m_3^2)^5} \Big[m_1^{12} + m_2^2 m_1^{10}
+ m_3^2 m_1^{10} - 15 m_2^2 m_3^2 m_1^8 
$$

$$
+ 10 m_2^2 m_3^4 m_1^6 + 10 m_2^4 m_3^2 m_1^6 - 5 m_2^2 m_3^6 m_1^4 
+ 5 m_2^4 m_3^4 m_1^4 - 5 m_2^6 m_3^2 m_1^4 
$$

$$
+ m_2^2 m_3^8 m_1^2 - 4 m_2^4 m_3^6 m_1^2 - 4 m_2^6 m_3^4 m_1^2 
+ m_2^8 m_3^2 m_1^2 + m_2^4 m_3^8 + m_2^6 m_3^6 + m_2^8 m_3^4 \Big]
$$

$$
+\,  \frac{1}{6 (m_1^2 - m_2^2)^4 (m_1^2 - m_3^2)^4} \Big[14 m_1^{10}
- 5 m_2^2 m_1^8 - 5 m_3^2 m_1^8 + 4 m_2^4 m_1^6 + 4 m_3^4 m_1^6
$$

$$
- 60 m_2^2 m_3^2 m_1^6 - m_2^6 m_1^4 - m_3^6 m_1^4 
+ 39 m_2^2 m_3^4 m_1^4 + 39 m_2^4 m_3^2 m_1^4 - 10 m_2^2 m_3^6 m_1^2 
$$

$$
- 6 m_2^4 m_3^4 m_1^2 - 10 m_2^6 m_3^2 m_1^2 - m_2^4 m_3^6 
- m_2^6 m_3^4 \Big] \Bigg\} + {\cal O} (r^6) \, ;
$$

$$
C_0 (0, q^2, 0;  m^2, m^2, m^2) = - \frac{1}{2 m^2} \Bigg[
1 + \frac{q^2}{12 m^2} +  \frac{q^4}{90 m^4} + {\cal O} (q^6) \Bigg] \ ;
$$

$$
C_0 (r^2, 0, r^2;  m^2, m^2, m^2) = - \frac{1}{2 m^2} \Bigg[
1 + \frac{r^2}{6 m^2} +  \frac{r^4}{30 m^4} + {\cal O} (r^6) \Bigg] \ .  
\\ \eqno{(B.3)}
$$

\end{document}